\documentclass[11pt]{article}
\usepackage[utf8]{inputenc}
\usepackage{lipsum}
\usepackage{amsfonts}
\usepackage{amssymb}
\usepackage{hyperref}
\usepackage{enumitem}
\usepackage{float}
\usepackage{graphicx}
\usepackage{caption} 
\usepackage{calc}
\usepackage{fancyhdr}
\usepackage{tcolorbox}
\usepackage{hyperref}
\usepackage{wrapfig}
\usepackage{graphicx} % Required for inserting images
\usepackage{amssymb}
\usepackage{amsmath}
\usepackage{multicol}
\usepackage{braket}
\usepackage{abstract}
\usepackage{multicol}
\usepackage{xcolor}
\usepackage{appendix}
\usepackage{graphicx}
\usepackage{subcaption}
\title{\normalsize\bf BMS transformed Quantum String Dynamics near a Black Hole}
\author{\normalsize {\sc Nihar Ranjan Ghosh\footnote{\tt g.nihar@iitg.ac.in}\ \ and Malay K. Nandy\footnote{\tt mknandy@iitg.ac.in \rm (Corresponding Author)}}\\
\normalsize \em Department of Physics, Indian Institute of Technology Guwahati\\
\normalsize \em Guwahati 781 039, India}
\date{(April 27, 2026)}
\begin{document}

\maketitle
\begin{abstract}
Asymptotic symmetries are expected to leave subtle but physically meaningful imprints on quantum probes of gravity, yet their manifestation in near-horizon dynamics remains incompletely understood. We examine this question for a closed bosonic string propagating in the near-horizon geometry of a five-dimensional Schwarzschild black hole subjected to a generalized Bondi-van der Burg-Metzner-Sachs (BMS) supertranslation. The extended nature of the string makes it especially sensitive to the resulting anisotropic geometric distortions, and this sensitivity appears most clearly in the angular sector of the worldsheet dynamics. Under the gauge and falloff conditions adopted here, the temporal and radial sectors remain unaffected by the supertranslation, while the angular deformation breaks the original 
SO$(4)$ symmetry of the background. The radial equation is governed by modified Bessel modes, with a nonvanishing radial conserved current, indicating transport-like propagation. Radial squeezing driven by gravity and anisotropic angular spreading induced by supertranslations provide a dynamical realization of string spreading near the horizon. Thus this analysis demonstrates that probe string dynamics encodes nontrivial signatures of BMS-induced deformations, providing a dynamical probe of symmetry structures in higher-dimensional black hole spacetimes.
\end{abstract}
\tableofcontents
\section{Introduction}
The behavior of matter near a black hole horizon remains a central problem in theoretical physics. Quantum effects in this regime are not only of intrinsic interest, but are also closely tied to some of the deepest questions in gravity, including the microscopic origin of black hole entropy and the information loss paradox. Understanding the quantum dynamics of matter as it approaches the horizon may therefore provide valuable insight into the interplay between gravity and quantum mechanics.

String theory, as a leading candidate for a consistent theory of quantum gravity, offers a natural framework in which to explore these issues. Because strings are extended objects, their dynamics probe geometric structure more sensitively than point particles, making them especially well suited for studying near-horizon physics and anisotropic deformations of the background.

The dynamics of particles and strings in curved spacetimes have long attracted attention, both for their conceptual importance and for their relevance to astrophysical phenomena. Such dynamics influence the structure and stability of accretion disks, gravitational lensing, and the emission of gravitational waves, and therefore connect fundamental theory with observational astrophysics \cite{PhysRevD.83.124015},\cite{Yi_2020},\cite{PhysRevD.83.044053},\cite{hanan2007chaotic},\cite{hosur2016chaos},\cite{gur2016chaos},\cite{PhysRevD.95.066014},\cite{PhysRevD.95.084037},\cite{panis2019determination},\cite{PhysRevD.93.084012}. A number of studies have examined particle motion in black hole spacetimes. For example, Panis et al.\cite{panis2019determination} showed that particle motion in Schwarzschild spacetime is periodic, whereas in Kerr-Newman spacetime it becomes quasi-periodic. By contrast, the motion of charged particles in magnetized or charged black hole backgrounds is often chaotic \cite{Yi_2020},\cite{PhysRevD.83.044053},\cite{hanan2007chaotic},\cite{PhysRevD.95.084037},\cite{PhysRevD.99.044012}.

String propagation in curved backgrounds has also been studied extensively. Early work by de Vega and Sanchez \cite{DEVEGA1987320} treated the spacetime geometry exactly while solving the string equations of motion and constraints perturbatively, leading to the mass spectrum and vertex operators in de Sitter space. Semiclassical quantization of oscillatory circular strings in de Sitter and anti-de Sitter backgrounds was carried out in \cite{PhysRevD.51.6917}, where it was shown that mass levels in AdS are evenly spaced, whereas in de Sitter they decay through tunneling.

Classical string propagation in and near black hole horizons has likewise been explored in detail. In \cite{PhysRevD.53.3296}, the authors studied string behavior near the singularity and horizon, finding divergent momentum and size at the singularity and radiation-like behavior in the near-singularity regime. String motion in Kerr-Newman backgrounds was investigated in \cite{FROLOV1989255}, where time-dependent configurations were mapped to geodesic motion in an effective three-dimensional metric, and infinitely long stationary string solutions were obtained outside the event horizon. String instabilities were identified in Schwarzschild and Reissner-Nordstr\"om spacetimes in \cite{PhysRevD.47.4498}, where the time component remained stable while the radial component was unstable under small fluctuations. The behavior of ring-shaped strings in FRW and Schwarzschild backgrounds was studied in \cite{PhysRevD.49.763}, showing that initial conditions determine whether a string is absorbed or escapes. In \cite{PhysRevD.51.6929}, elliptic-function solutions for open and closed strings in curved spacetimes were derived, and it was shown that straight strings falling into Schwarzschild black holes grow without bound near the singularity. Circular strings in Schwarzschild, Reissner-Nordstr\"om, and de Sitter spacetimes were examined in \cite{PhysRevD.50.2623}, revealing divergent radial perturbations as $r\to0$ while angular perturbations remain bounded.

A particularly intriguing idea is that strings may spread over the horizon, forming a membrane-like structure known as the stretched horizon \cite{thorne1986membrane},\cite{hooft1985quantum},\cite{hooft1990black},\cite{GtHooft_1991},\cite{PhysRevLett.71.2367},\cite{PhysRevD.48.3743},\cite{PhysRevD.50.2725}. ’t Hooft \cite{hooft1985quantum},\cite{hooft1990black} proposed that the black hole horizon may be described by an operator algebra on a two-dimensional surface, closely resembling the worldsheet description of strings. In this picture, the black hole is treated as a classical statistical system on its membrane, while the string remains a quantum object, and the two descriptions may be related through Wick rotation \cite{GtHooft_1991}.

On the other hand, the Bondi-van der Burg-Metzner-Sachs (BMS) symmetry group and the associated soft-hair proposal have emerged as promising frameworks in the study of black hole information. The BMS symmetry group \cite{bondi1962gravitational,sachs1962gravitational} is the infinite-dimensional symmetry group of asymptotically flat spacetimes, extending the finite-dimensional Poincare group by an infinite set of generators known as supertranslations. These transformations may be interpreted as angle-dependent translations near null infinity and enrich the phase space structure of general relativity by distinguishing between spacetime configurations that are otherwise equivalent under Poincar'e symmetry \cite{ghosh2025asymptotic}. The emergence of this symmetry group was one of the unexpected outcomes of the original analysis of asymptotically flat spacetimes, and it has since been tied to the infrared triangle \cite{strominger2018lecturesinfraredstructuregravity}, whose three corners comprise asymptotic symmetries, soft theorems, and gravitational memory effects.

Similar developments have occurred in asymptotically anti-de Sitter spacetimes. In particular, Brown and Henneaux \cite{brown1986central} showed that the asymptotic symmetry algebra of $\mathrm{AdS}_3$ gravity consists of two copies of the Virasoro algebra. These results highlight the broader universality and physical significance of asymptotic symmetry analysis \cite{CADONI1999165, hotta1998asymptoticisometrydimensionalantide, henneaux1985asymptotically, Comp_re_2016}. It may aso be noted that physical consraints  on the infinite-dimensional BMS symmetry group has been recently studied in Ref.~{ghosh2026cons}.

Within the infrared triangle, gravitational memory provides a classical and observable manifestation of BMS symmetry. First identified in the linearized regime by Zel'dovich and Polnarev \cite{Zeldovich:1974gvh} and later extended to the nonlinear regime by Christodoulou and others \cite{Christodoulou, Braginsky:1985vlg, Braginsky:1987kwo, PhysRevD.44.R2945, PhysRevD.46.4304, PhysRevD.45.520, Favata:2010zu, Tolish:2014bka, Tolish:2014oda, Winicour:2014ska}, the memory effect describes the permanent displacement of freely falling test masses due to gravitational-wave propagation. This permanent displacement is directly linked to BMS supertranslations, since memory encodes the radiative history of spacetime \cite{Strominger:2014pwa, strominger2018lecturesinfraredstructuregravity, PhysRevD.92.084057, Flanagan:2015pxa, Pasterski:2015tva}. The notion has since been enlarged to include spin and center-of-mass memory effects \cite{Pasterski:2015zua, PhysRevD.98.064032, Compere:2016jwb}, as well as electromagnetic analogues \cite{Bieri:2013hqa, susskind2019electromagneticmemory}. In the black hole context, Donnay et al.\cite{PhysRevLett.116.091101} introduced black hole memory and showed that transient radiation can induce permanent deformations of the near-horizon geometry, providing a route from memory effects to the emergence of soft hair on horizons. Related features in extremal black holes have also been identified \cite{PhysRevD.102.044041}.

The third corner of the infrared triangle, soft theorems, has its origin in early studies of infrared behavior in quantum electrodynamics, notably in the work of Bloch and Nordsieck \cite{PhysRev.52.54}, Low \cite{Low:1954kd, Low:1958sn}, Gell-Mann and Goldberger \cite{Gell-Mann:1954wra}, and Yennie et al.\ \cite{Yennie:1961ad}. The gravitational version of soft theorems was developed by Weinberg \cite{Weinberg:1965nx}, who showed that the emission of low-energy gravitons in scattering processes obeys universal behavior independent of the detailed microscopic dynamics. These universal properties are now understood to correspond to Ward identities associated with asymptotic symmetries, thereby establishing a direct link between soft theorems and BMS charges \cite{Pasterski:2015tva}. In this sense, soft theorems provide a quantum-field-theoretic realization of the infinite-dimensional symmetry structure uncovered by BMS analysis and form a central ingredient in the soft-hair picture of black holes.

The soft-hair proposal has attracted considerable attention because of its possible relevance to the black hole information loss paradox \cite{PhysRevD.14.2460}. Hawking, Perry, and Strominger \cite{PhysRevLett.116.231301} argued that the infinite number of conserved BMS charges corresponds to infinitely many zero-energy excitations, or soft hair, associated with the black hole horizon. In principle, these soft degrees of freedom could encode information about black hole microstates and thereby provide a mechanism for information retention during Hawking evaporation \cite{Hawking:1975vcx, Hawking:2016sgy, haco2018black, PhysRevD.96.084032, Chu_2018, PhysRevD.108.044034}. This proposal modifies the classical no-hair theorem \cite{Israel:1967za, Israel.164.1776, PhysRevLett.26.331}, which states that black holes are characterized only by a finite set of global charges, and suggests an infinite-dimensional structure not captured by the traditional picture \cite{strominger2017blackholeinformationrevisited}. Such ideas have important implications for black hole entropy and unitarity in quantum gravity \cite{haco2019kerrnewmanblackholeentropy, PhysRevD.103.126020}. An alternative but complementary viewpoint interprets soft hair as edge modes, namely boundary-localized degrees of freedom arising from large gauge transformations that act nontrivially at spacetime boundaries. These edge modes are essential for maintaining gauge invariance on subregions and have been widely studied in gauge theories and gravity \cite{Donnelly:2014fua, Donnelly:2015hxa, Harlow:2015lma, Harlow:2016vwg, Maldacena:2016upp}.

%Since strings are fundamental extended objects and the BMS group characterizes the asymptotic symmetries of flat spacetimes, such an analysis may provide a useful bridge between near-horizon string dynamics and asymptotic gravitational data.

Motivated by these developments, it is natural to ask how a large gauge transformation, such as a BMS transformation, affects the dynamics of a string propagating in a black hole background.  Specifically, one may ask how the string wavefunction responds when the background metric is deformed by a large gauge transformation, and whether the resulting dynamics encode any physically meaningful imprint of the BMS transformation.
Since strings have extended structure,
sensitive to angular distortions and
couples to geometry differently via induced metric, we consider the quantum dynamics of a closed string near the horizon of a $5$D Schwarzschild black hole, with the background geometry subjected to a generalized BMS symmetry transformation in $5$D.
Our analysis is based on the Polyakov action in the conformal gauge, along with the background geometry treated purely through its spacetime metric with the exclusion of the Ramond-Ramond $(RR)$ fields and worldsheet fermions.

In the present work, we therefore consider a bosonic string propagating in an asymptotically flat, BMS-deformed five-dimensional Schwarzschild background. In this limit, string dynamics is well approximated by the low-energy effective description, wherein the background geometry satisfies the vacuum Einstein equations, corresponding to the massless sector of string theory \cite{Polchinski:1998rq},\cite{Green:1987sp}. This stands in contrast to backgrounds such as $AdS_5\times S^5$
, which arise in type IIB string theory supported by Ramond–Ramond five-form flux, and for which a consistent worldsheet description requires the Green–Schwarz or pure spinor formalism \cite{berkovits2000super},\cite{callan2003quantizing},\cite{metsaev1998type}. Quantization in such RR-supported backgrounds is technically involved and is typically tractable only in special limits, such as the Penrose (pp-wave) or BMN regime \cite{berenstein2002strings}. By contrast, in the present asymptotically flat setting, we do not attempt a full fermionic formulation and instead focus on the bosonic sector of the theory. The resulting framework provides a minimal but controlled setting to probe how BMS-induced deformations influence string dynamics, capturing in particular qualitative features such as angular spreading near the horizon.

The rest of the paper is organised as follows. In Section \ref{model}, we introduce the BMS transformation and thereby construct the modified Polyakov action for a string propagating in the background of a $5$D Schwarzschild black hole. Hence we find the constraints that generates the Hamiltonian leading to the Schr\"odinger equation for the string. We find its general solution in Section \ref{solution} using suitable techniques with appropriate approximation for the near-horizon behaviour. Section \ref{l=2} visualises the wavefunction in the $l=2$ hyperspherical mode where we also plot the angular and radial sectors of the wave functions. Finally, we conclude the paper with a discussion in Section \ref{disc}.

\section{BMS transformation and string Schr\"odinger equation}\label{model}
We  investigate the dynamics of a string  under BMS transformation in the curved background of a $5$D Schwarzschild black hole. The background metric has the form 
\begin{equation}
    \label{sch}
    ds^2\Big|_{\bar{g}_{\mu\nu}}=-fdt^2+f^{-1}dr^2+r^2(d\theta^2+\sin^2\theta d\chi^2+\cos^2\theta d\phi^2)~,
\end{equation}
with $f=1-2M/r^2$. The string coordinates will be represented by $X^\mu=(t,r,\theta,\chi,\phi)$, with  $t=t(\tau),~ r=r(\tau), ~ \theta=\theta(\tau),~\phi=\phi(\tau)$ and $\chi=\chi(\sigma)$, where $\tau$ and $\sigma$ parametrize the string world sheet. The Greek alphabets span as $\mu,\nu=0,1,\dots 5$ and Latin alphabets as $a,b=0,1$ for  $\tau$ and $\sigma$ respectively.

It may be noted that we work within a truncated sector of the theory, which is analogous to the
mini-superspace approximation, where we restrict our attention to the class of string configurations
depending only on the worldsheet time parameter $\tau$, with a constant winding number $k$ along
one angular direction $\chi=k\sigma$ . This ultimately reduces the infinite-dimensional phase
space of the full string theory to a finite dimensional system of effective degrees of freedom, which simplifies the dynamics but preserve the essential features of the model.

For convenience, we express the above metric \ref{sch} in retarded coordinates as
\begin{equation}
    \label{metric in u}
     ds^2\Big|_{\bar{g}_{\mu\nu}}=-fdu^2-2dudr+r^2(d\theta^2+\sin^2\theta d\chi^2+\cos^2\theta d\phi^2)~,
\end{equation}
and consider the class of BMS transformations with gauge conditions
\begin{equation}
    \label{BMS}
    \mathcal{L}_\eta \bar{g}_{uu}=\mathcal{L}_\eta \bar{g}_{rA}=\mathcal{L}_\eta \bar{g}_{rr}=\mathcal{L}_\eta \bar{g}_{ur}=0~,
\end{equation}
with falloff conditions $\eta^u \sim \mathcal{O}(1);~\eta^r\sim\mathcal{O}(r^{-2}),~\eta^\theta, \eta^\phi\sim \mathcal{O}(r^{-1})$. 

The vector field satisfying the above conditions is given by
\begin{equation}
    \label{eta}
    \eta=F\partial_u+\left(\frac{2M}{r^2}-1 \right)F\partial_r-\frac{1}{r}\left[ (\partial_\theta F )\partial_\theta+\frac{\partial_\chi F}{\sin^2\theta}\partial_\chi +\frac{\partial_\phi F}{\cos^2\theta} \partial_\phi     \right]~.
\end{equation}
This gauge and the falloff conditions are not exactly the $4$D BMS conditions, but rather a generalization of the BMS conditions to $5$D. For $F=\text{constant}$, and in the limit $r\to r_h$ (black hole horizon), the vector field $\eta$ results in a time translation, whereas  in the asymptotic limit $r\to\infty$,  a combination of time and radial translations is obtained. Moreover, the diffeomorphic vector field $\eta$ keeps the energy-momentum finite, preserves the horizon structure, and preserves asymptotic flatness, as will be clear  below from the transformed metric.

Under the action of the above diffeomorphic vector field, the transformed metric $g_{\mu\nu}=\bar{g}_{\mu\nu}+\mathcal{L}_{\eta}\bar{g}_{\mu\nu}$ gives the line element as
\begin{equation}
    \label{new met}
   \begin{split}
        ds^2\Big|_{g_{\mu\nu}}=&\left[\bar{g}_{\mu\nu}+\mathcal{L}_{\eta}\bar{g}_{\mu\nu}\right]dx^\mu dx^\nu=-f du^2-2dudr+r^2(d\theta^2+\sin^2\theta d\chi^2+\cos^2\theta d\phi^2)\\
        &+r(C_{\theta\theta}d\theta^2+C_{\chi\chi}d\chi^2+C_{\phi\phi}d\phi^2)+2(g_{\theta\chi}d\theta d\chi+g_{\theta\phi}d\theta d\phi+g_{\phi\chi}d\phi d\chi)\\
        =&-fdt^2+f^{-1}dr^2+r^2(d\theta^2+\sin^2\theta d\chi^2+\cos^2\theta d\phi^2)\\
        &+r(C_{\theta\theta}d\theta^2+C_{\chi\chi}d\chi^2+C_{\phi\phi}d\phi^2)+2(g_{\theta\chi}d\theta d\chi+g_{\theta\phi}d\theta d\phi+g_{\phi\chi}d\phi d\chi)~,
   \end{split}
\end{equation}
with 
\begin{equation}
    \label{coeff}
    \begin{split}
        C_{\theta\theta}&=-2(\partial_\theta^2F+F),\\
        C_{\chi\chi}&=-2(\sin^2\theta F+\sin\theta\cos\theta\partial_\theta F+\partial_\chi^2F),\\
        C_{\phi\phi}&=-2(\cos^2\theta F-\sin\theta\cos\theta\partial_\theta F+\partial_\phi^2F),\\
        g_{\theta\chi}&=-r\partial_\chi\left[\partial_\theta F+\sin^2\theta \partial_\theta\left( \frac{F}{\sin^2\theta} \right)    \right],\\
        g_{\theta\phi}&=-r\partial_\phi\left[\partial_\theta F+\cos^2\theta \partial_\theta\left( \frac{F}{\cos^2\theta} \right)    \right],\\
        g_{\phi\chi}&=-2r\partial_\phi\partial_\chi F~.
    \end{split}
\end{equation}

As a direct consequence of the particular gauge and falloff conditions chosen, the temporal and radial parts of the metric remain unchanged in all orders of $r$. Furthermore, it is evident from the transformed metric that the BMS induced deformation introduces angular dependent shear on the three sphere ($S_3$), breaking the original $SO(4)$ symmetry and leading to an anisotropic angular geometry. 

Having obtained the transformed metric, the corresponding $5$D Polyakov action, transformed by the BMS transformation, is given by
\begin{equation}
    \label{Lagarangian}
    \begin{split}
        \mathcal{S}&=\frac{-1}{4\pi\alpha'}\int d\tau d\sigma\sqrt{-h}h^{ab}g_{\mu\nu}\partial_aX^\mu\partial_bX^\nu\\
        &=\frac{1}{4\pi\alpha'}\int d\tau d\sigma\Big[   -f\left( \frac{dt}{d\tau} \right)^2+f^{-1}\left( \frac{dr}{d\tau} \right)^2+(r^2+rC_{\theta\theta})\left( \frac{d\theta}{d\tau} \right)^2\\&+(r^2\cos^2\theta+rC_{\phi\phi})\left( \frac{d\phi}{d\tau} \right)^2+2g_{\theta\phi}\left( \frac{d\theta}{d\tau} \right)\left( \frac{d\phi}{d\tau} \right)+k^2(r^2\sin^2\theta+rC_{\chi\chi})\Big]~,
    \end{split}
\end{equation}
where in the last equality we have considered the conformal gauge in which the $2$D world sheet metric is flat $h_{ab}=\text{diag}(-1,1)$. Importantly, because of the breaking of $SO(4)$ symmetry, the BMS induced deformation of the angular sector enters the worldsheet action through the perturbed target space metric. This manifests via an angle dependent anisotropic contribution to the effective potential  after mode decomposition. 

The explicit form of the Lagrangian is needed to compute the Hamilton's equations and to find the canonical momenta corresponding to the string coordinates.

Now, the first string constraint equation
\begin{equation}
    \label{constraint 1}
g_{\mu\nu}\left[\frac{\partial X^\mu}{\partial\tau}\frac{\partial X^\nu}{\partial\tau}+\frac{\partial X^\mu}{\partial\sigma}\frac{\partial X^\nu}{\partial\sigma}    \right]=0
\end{equation}
leads to
\begin{equation}
    \label{constraint 11}
    \begin{split}
        -f\left( \frac{dt}{d\tau} \right)^2+f^{-1}\left( \frac{dr}{d\tau} \right)^2&+(r^2+rC_{\theta\theta})\left( \frac{d\theta}{d\tau} \right)^2+(r^2\cos^2\theta+rC_{\phi\phi})\left( \frac{d\phi}{d\tau} \right)^2\\&+2g_{\theta\phi}\left( \frac{d\theta}{d\tau} \right)\left( \frac{d\phi}{d\tau} \right)+k^2(r^2\sin^2\theta+rC_{\chi\chi})=0~.
    \end{split}
\end{equation}
Moreover, the second constraint equation
\begin{equation}
    \label{constraint 2}
    g_{\mu\nu}\frac{\partial X^\mu}{d\tau}\frac{\partial X^\nu}{d\sigma}=0~~~\text{gives}~~~g_{\theta\chi}\frac{d\theta}{d\tau}+g_{\phi\chi}\frac{d\phi}{d\tau}=0~,
\end{equation}
provided $k\neq0$. 

The first constraint \ref{constraint 11} determines the equation of motion of the string and the second constraint \ref{constraint 2} restricts the angular sector of the string wave function.

For simplification of the dynamical equation, we choose $g_{\theta\phi}=0$. This gauge choice leads to $2\partial_\theta F+2F\tan\theta=\mathcal{F}(\theta,\chi)$, where $\mathcal{F}(\theta,\chi)$ is an arbitrary function,  leaving us with enough gauge freedom without losing generality in the rest of the analysis.

Substitution of $g_{\theta\phi}=0$ in equation \ref{Lagarangian} simplifies the Hamilton's equations of motion to
\begin{equation}
    \label{hamiltons constraint}
    \begin{split}
        p_t&=\frac{\partial\mathcal{L}}{\partial \dot{t}}=\frac{-1}{2\pi\alpha'}f\dot{t},\\
        p_r&=\frac{\partial\mathcal{L}}{\partial \dot{r}}=\frac{1}{2\pi\alpha' f}\dot{t},\\
        p_\theta&=\frac{\partial\mathcal{L}}{\partial \dot{\theta}}=\frac{1}{2\pi\alpha'}(r^2+rC_{\theta\theta})\dot{\theta},\\
        p_\phi&=\frac{\partial\mathcal{L}}{\partial \dot{\phi}}=\frac{1}{2\pi\alpha'}(r^2\cos^2\theta+rC_{\phi\phi})\dot{\phi},\\
        p_\chi&=\frac{\partial\mathcal{L}}{\partial \dot{\chi}}=0,
    \end{split}
\end{equation}
giving
\begin{equation}
    \label{momentum}
    \begin{split}
        \dot{t}&=-\frac{2\pi\alpha'}{f}p_t,\\
        \dot{r}&=2\pi\alpha' f p_r,\\
        \dot{\theta}&=\frac{2\pi\alpha'}{r^2+rC_{\theta\theta}}p_\theta,\\
        \dot{\phi}&=\frac{2\pi\alpha'}{r^2\cos^2\theta+rC_{\phi\phi}}p_\phi~.\\
    \end{split}
\end{equation}

Consequently, the Hamiltonian $H=p_t\dot{t}+p_r\dot{r}+p_\theta\dot{\theta}+p_\phi\dot{\phi}-\mathcal{L}$ is given by

\begin{equation}
    \label{hamiltonian}
    \begin{split}
        H=\pi\alpha'\left[ -\frac{1}{f}p_t^2+fp_r^2+\frac{1}{r^2+rC_{\theta\theta}}p_\theta^2+\frac{1}{r^2\cos^2\theta+rC_{\phi\phi}}p_\phi^2 \right]+\frac{k^2}{4\pi\alpha'}(r^2\sin^2\theta+r C_{\chi\chi})~.
    \end{split}
\end{equation}
With the substitution of equations \ref{hamiltons constraint} and \ref{momentum} in \ref{constraint 11}, one obtains
\begin{equation}
    \label{operator equ}
    H=0~.
\end{equation}

As mentioned earlier, in the considered model of mini superspace like configuration, variation of the Polyakov action with respect to the worldsheet metric yields the Hamiltonian constraint $H = 0$, that reflects the residual gauge invariance for specific worldsheet configuration. We obtain the dynamical equation for the string by promoting  this to a quantum constraint equation of the form $\hat{H}\Psi(t,r,\theta,\chi,\phi) = 0$.

Importantly, in the full string theory, physical states are subject to the full set of Virasoro constraints
$T_{ab} = 0$, that are implemented mode-by-mode via the full tower of Virasoro generators $L_n$ (e.g.,
$\hat{L}_n\Psi = 0$), which results in a highly detailed structure of the physical state space. On the other hand, our approach does not attempt to capture such full structure. Instead, the quantum constraint
$\hat{H}\Psi(t,r,\theta,\chi,\phi) = 0$ effectively reflects the zero-mode sector of the Virasoro constraints, providing a tractable framework to explore quantum aspects of string dynamics in this curved black
hole spacetime.

Thus, promoting equation \ref{operator equ} to an operator $\hat{H}$ with the transformation $p_\mu\to \hat{p}_\mu= -i\hbar\frac{\partial}{\partial X^\mu}$, we have 
\begin{equation}
\label{main equation}
\Bigg[\hbar^2\left\{-\partial_t^2+f^2\partial_r^2+\frac{f}{r^2+rC_{\theta\theta}}\partial_\theta^2+\frac{f}{r^2\cos^2\theta+rC_{\phi\phi}}\partial_\phi^2 \right\} -\frac{
k^2f}{4\pi^2\alpha'^2}(r^2\sin^2\theta+rC_{\chi\chi})\Bigg]\Psi(t,r,\theta,\chi,\phi)=0,
\end{equation}
where $\Psi(t,r,\theta,\chi,\phi)$ represents the wave function of the string, living in the $5$D background spacetime with the metric \ref{new met}. Solution of equation \ref{main equation} will yield the quantum mechanical behaviour of the closed string in the reduced setting considered here.

\section{Analytical solution for the string wavefunction}
\label{solution}
\begin{figure}[h!]
    \centering
    \begin{subfigure}{0.8\textwidth}
        \centering
        \includegraphics[width=\linewidth]{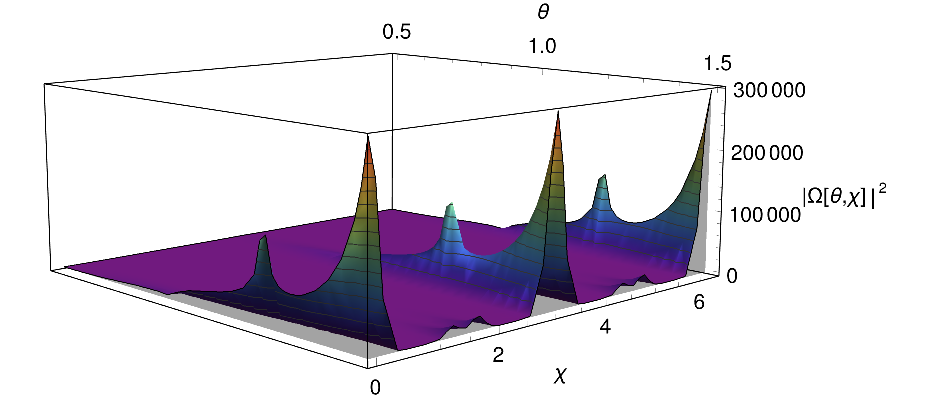}
        \caption{Right-moving probability density $|\Omega(\theta,\chi)|^2$.}
        \label{eta_squared_B=0}
    \end{subfigure}
    \begin{subfigure}{0.8\textwidth}
        \centering
        \includegraphics[width=\linewidth]{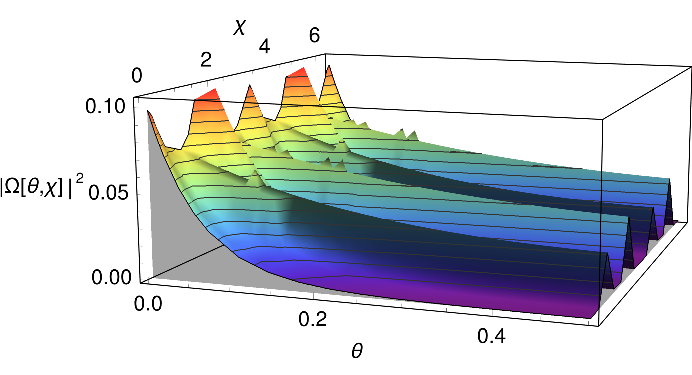}
        \caption{Left-moving probability density $|\Omega(\theta,\chi)|^2$.}
        \label{eta_squared_A=0}
    \end{subfigure}
    \caption{String probability density  $|\Omega(\theta,\chi)|^2$ in the angular sector for the right-moving  and left-moving modes.}
    \label{3D eta squared}
\end{figure}
The temporal part of the wave function can be obtained with the substitution $\Psi(t,r,\theta,\chi,\phi)=T(t)\psi(r,\theta,\chi,\phi)$ in equation \ref{main equation}, leading to 
\begin{equation}
    \label{time part eqn}
     \frac{\hbar^2}{T}\frac{d^2T}{dt^2}=-\epsilon^2~,
\end{equation}
and 
\begin{equation}
    \label{time minus remain}
    \frac{1}{\psi}\Bigg[\hbar^2\left\{f^2\partial_r^2+\frac{f}{r^2+rC_{\theta\theta}}\partial_\theta^2+\frac{f}{r^2\cos^2\theta+rC_{\phi\phi}}\partial_\phi^2 \right\} -\frac{
k^2f}{4\pi^2\alpha'^2}(r^2\sin^2\theta+rC_{\chi\chi})\Bigg]\psi=-\epsilon^2~,
\end{equation}
with $\epsilon^2\geq0$. The solution of equation \ref{time part eqn} is given by 
\begin{equation}
    \label{time sol}
    T(t)=C_1\exp{\left[\frac{i\epsilon t}{\hbar}\right]}+C_2\exp{\left[-\frac{i\epsilon t}{\hbar}\right]}~,
\end{equation}
where $C_1$ and $C_2$ are two arbitrary integration constants. As a consequence of the gauge choice in equation \ref{BMS}, the temporal part of the metric remains unchanged, resulting in the temporal part of the wavefunction a linear combination of stationary state solutions.

Moreover, with the substitution of $\psi(r,\theta,\chi,\phi)=\tilde{\psi}(r,\theta,\chi)\Phi(\phi)$ in equation \ref{time minus remain}, we have
\begin{equation}
    \label{phi equation}
    -\frac{\hbar^2}{\Phi}\frac{d^2\Phi}{d\phi^2}=\lambda^2~
\end{equation}
and 
\begin{equation}
    \label{time+phi minus remain}
    \frac{\hbar^2}{\tilde{\psi}}(r^2\cos^2\theta+rC_{\phi\phi})\Bigg[f\partial_r^2+\frac{1}{r^2+rC_{\theta\theta}}\partial_\theta^2 -\frac{
k^2}{4\pi^2\alpha'^2\hbar^2}(r^2\sin^2\theta+rC_{\chi\chi})+\frac{\epsilon^2}{\hbar^2 f }\Bigg]\tilde{\psi}=\lambda^2~.
\end{equation}
Solution of equation \ref{phi equation} turns out to be
\begin{equation}
    \label{phi solution}
\Phi(\phi)=C_3\exp{\left[\frac{i\lambda\phi}{\hbar}\right]}+C_4\exp{\left[-\frac{i\lambda\phi}{\hbar}\right]}~,
\end{equation}
with $\lambda^2\geq0$ the separation constant, and $C_3$, $C_4$ are two arbitrary integration constants. Just like  the time part of the wavefunction, the azimuthal part of wavefunction also results in a linear combination of plane wave solutions because of the gauge choice $F=F(\theta,\chi)$. Since $\phi$ is a periodic coordinate with period $2\pi$, the wavefunction must satisfy the single value condition $\Phi(\phi+2\pi)=\Phi(\phi)$. This imposes the constraint $e^{i2\pi \lambda/\hbar}=1$, which leads to the quantization condition $\lambda = n\hbar$, where $n \in \mathbb{Z}$. Consequently, the separation constant takes discrete values $\lambda^2 = n^2 \hbar^2$. Equivalently, the angular dependence may be expressed in terms of trigonometric functions,
$\Phi(\phi)=A_1\cos(n\phi)+B_1\sin(n\phi),$
corresponding to eigenfunctions of the azimuthal angular momentum operator $-i\hbar\,\partial_\phi$.

Thus the methodology to find the solution of temporal and azimuthal parts of the wavefunction is straightforward. However, the nonlinear coupling between the radial coordinate $r$ and the remaining angular coordinates $\theta $ and $ \chi$, together with the arbitrariness of the supertranslation parameter $F(\theta,\chi)$, makes the analytical solution to equation \ref{time+phi minus remain} nearly impossible. Nevertheless, one can solve equation \ref{time+phi minus remain} by expanding about a small perturbation parameter $x=r-r_0$, with $r_0=\sqrt{2M}$ the horizon of the black hole. 

Since we are interested in obtaining the nature of the wave function near the horizon subject to the BMS symmetry transformation, it would be legitimate to capture the non-linear nature of the system up to the contribution $\mathcal{O}(x^2)$ in the radial part. Thus, substituting $\tilde{\psi}(r,\theta,\chi)=R(x)\Omega(\theta,\chi)$, equation \ref{time+phi minus remain} gives  
\begin{equation}
    \label{theta chi equn 2}
    \frac{\partial^2\Omega}{\partial\theta^2}-V(\theta,\chi)\Omega=0~,
\end{equation}
with
\begin{equation}
    \label{xi}
    V(\theta,\chi)=\frac{\lambda^2(C_{\theta\theta}+r_0)}{\hbar^2(C_{\phi\phi}+r_0\cos^2\theta)}+\frac{k^2r_0^2}{4\pi^2\alpha'^2\hbar^2}(C_{\chi\chi}+r_0\sin^2\theta)(C_{\theta\theta}+r_0)-M\beta(C_{\theta\theta}+r_0)~,
\end{equation}
and
\begin{equation}
    \label{r equation}
    x^2\frac{d^2R}{dx^2}+\left( \frac{M\epsilon^2}{2\hbar^2}-\frac{M\beta}{2}x \right)R=0~,
\end{equation}
where $\beta$ is the separation constant.
\begin{figure}[h!]
    \centering
    \begin{subfigure}{0.45\textwidth}
        \centering
        \includegraphics[width=\linewidth]{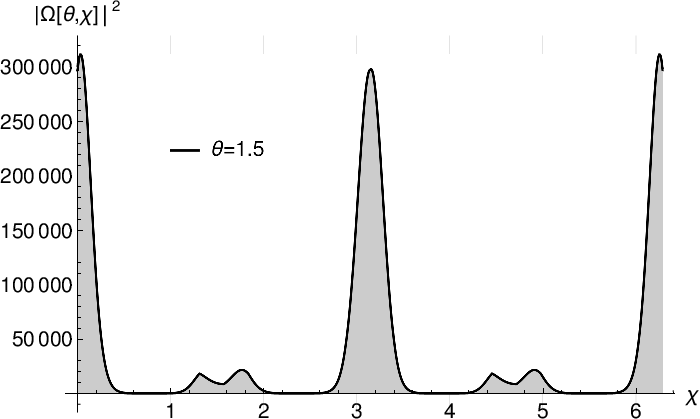}
        \caption{Periodicity of $|\Omega(\theta,\chi)|^2$ for the right-moving mode with $\theta=1.5$.}
        \label{constant_theta_eta_squared_B=0}
    \end{subfigure}
    \hfill
    \begin{subfigure}{0.45\textwidth}
        \centering
        \includegraphics[width=\linewidth]{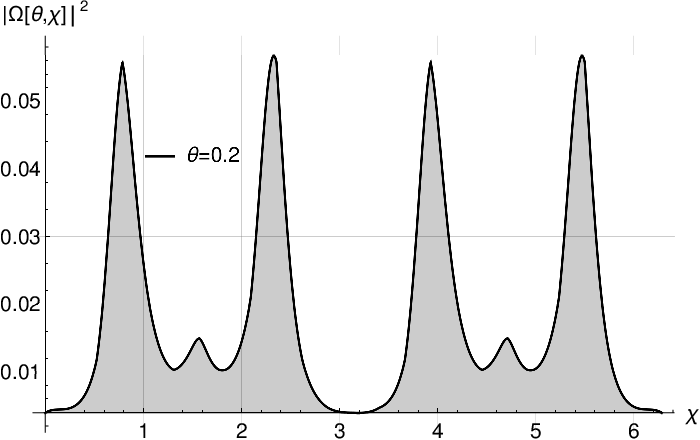}
        \caption{Periodicity of $|\Omega(\theta,\chi)|^2$ for the left-moving mode with $\theta=0.2$.}
        \label{constant_theta_eta_squared_A=0}
    \end{subfigure}
    \caption{Periodicity of the string probability density $|\Omega(\theta,\chi)|^2$ in the $\chi$-direction for right-moving and left-moving modes with constant values of $\theta$.}
    \label{chi periodicity}
\end{figure}
The solution of equation \ref{r equation} turns out to be
\begin{equation}
    \label{r solution}
    \begin{split}
        R(r)= \sqrt{\frac{ M\beta x}{2} } \Bigg[&  C_5~ i^{1-\sqrt{1-4 K}} ~  \Gamma \left(1-\sqrt{1-4 K}\right)~ I_{-\sqrt{1-4 K}}\left(\sqrt{2M\beta x}\right)\\
        &+ C_6 ~i^{1+\sqrt{1-4 K}} ~\Gamma \left(\sqrt{1-4 K}+1\right)~ I_{\sqrt{1-4 K}}\left(\sqrt{2M\beta x}\right)     \Bigg]\\
    \end{split}
\end{equation}

where $K=\frac{M\epsilon^2}{2\hbar^2}$ and $I_\nu(z)$ is the modified Bessel function of the first kind of order $\nu$. Since the radial dynamics is governed by an 
effective potential given by equation \ref{r equation}, it admits solutions in terms  
$I_{\pm \sqrt{1-4K}}$  which are exponential rather than oscillatory 
in the chosen variable $x$. The order of the Bessel function is real-valued for $K < 1/4$, leading to exponential solutions whereas it becomes imaginary for $K > 1/4$, 
leading to oscillatory behavior. 
The prefactor  $\sqrt{\frac{ 1}{2}M\beta x}$ in equation \ref{r solution} encodes the expected radial scaling, while 
the overall phases $i^{1\pm\sqrt{1-4K}}$ can be absorbed into the integration 
constants $C_5$ and $C_6$. 

Defining the radial flux density as
\begin{equation}
    \label{flux} 
    J^r=\frac{1}{2i}\left[R^*\partial_xR-R\partial_xR^* \right]
\end{equation}
one finds
\begin{equation}
    J^r =
\begin{cases}
\displaystyle \frac{m\beta}{2}\sqrt{1-4K}~~C_5C_6\sin{(\pi\sqrt{1-4K})}~~~~~~\text{if}~~K<\frac{1}{4}\\
\displaystyle\\
\displaystyle -\frac{m\beta}{2}\sqrt{4K-1}~~C_5C_6\sinh{(\pi\sqrt{4K-1})}~~\text{if}~~K>\frac{1}{4}
\end{cases}
\end{equation}
with $C_5,C_6\in \mathbb{R}$.

Thus, the value of the flux increases as the mass of the black hole increases and the radial flux is nonzero in general unless $K=\frac{M\epsilon^2}{2\hbar^2}=1/4$. The non-vanishing and real nature of the radial flux indicates that the 
solution carries a net radial current. This implies that the mode is 
non-localized and corresponds to a transport type
configuration rather than a trapped or purely evanescent mode.

Turning to the angular part, we employ the WKB approximation to equation \ref{theta chi equn 2} and obtain 
\begin{equation}
    \label{eta solution}
    \Omega(\theta,\chi)=\frac{1}{V^{1/4}}\left[A(\chi)\exp\left\{\int^\theta\sqrt{V(y,\chi)}dy\right\}+B(\chi)\exp\left\{-\int^\theta\sqrt{V(y,\chi)}dy\right\}  \right]
\end{equation}
up to leading order, with $V$ given by equation \ref{xi}. 

Consequently, equations \ref{time sol}, \ref{phi solution}, \ref{r solution}, \ref{eta solution} yield the complete wavefunction for the string as
\begin{equation}
    \label{wave function}
    \begin{split}
        &\Psi(t,x,\theta,\chi,\phi)=\Bigg[C_1\exp{\left\{\frac{i\epsilon t}{\hbar}\right\}}+C_2\exp{\left\{-\frac{i\epsilon t}{\hbar}\right\}}\Bigg]\Bigg[  C_3\exp{\left\{\frac{i\lambda\phi}{\hbar}\right\}}+C_4\exp{\left\{-\frac{i\lambda\phi}{\hbar}\right\}}  \Bigg]\\
        &\sqrt{\frac{x M\beta}{2} } \Bigg[ i^{1-\sqrt{1-4 K}}   C_5 \Gamma \left(1-\sqrt{1-4 K}\right) I_{-\sqrt{1-4 K}}\left(\sqrt{2Mx\beta}\right)+i^{1+\sqrt{1-4 K}} C_6 \Gamma \left(\sqrt{1-4 K}+1\right) I_{\sqrt{1-4 K}}\left(\sqrt{2Mx\beta}\right)     \Bigg]\\
        &\frac{1}{V^{1/4}}\left[A(\chi)\exp\left\{\int^\theta\sqrt{V(y,\chi)}dy\right\}+B(\chi)\exp\left\{-\int^\theta\sqrt{V(y,\chi)}dy\right\}  \right]
    \end{split}
\end{equation}
in the vicinity of the horizon.
 
It is important to note that the wave function in equation \ref{wave function} is in general valid for any transformation given by equation \ref{BMS} since we have not considered any particular form of the supertranslation parameter $F=F(\theta,\chi)$.

 As mentioned earlier, one of the main advantages of strings,  an extended object with nonzero angular distribution, is that they are very sensitive to the angular deformation of  spacetime. This will be clear in the next section where an explicit example is considered with a particular form of $F(\theta,\chi)$.

\section{Wave function visualization with $l=2$ hyperspherical mode} \label{l=2}
Having obtained the complete wave function of the string in the vicinity of the $5$D Schwarzschild black hole horizon, we now consider an explicit example of the supertranslation parameter $F(\theta,\chi)$.

To be specific, we choose the supertranslation parameter $F(\theta,\chi)$ to be the hyperspherical mode corresponding to $l=2, m=0 , p=2$, represented by $Y_{202}(\theta,\phi,\chi)$, where $l, m, p$ are the quantum numbers corresponding to the $\theta, \phi$ and $\chi$ modes respectively. 
Thus,
\begin{equation}
    \label{1st F}
    F(\theta,\chi)=a_1 \cos{2\theta}e^{i2\chi}
\end{equation}
\begin{figure}[h!]
    \centering
    \begin{subfigure}{0.45\textwidth}
        \centering
        \includegraphics[width=\linewidth]{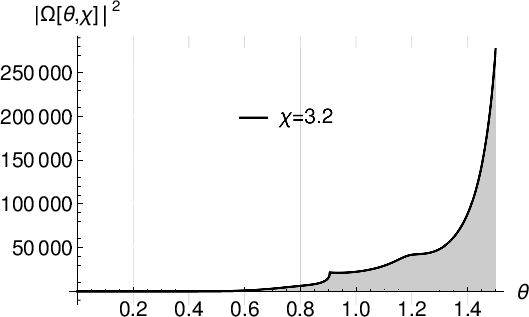}
        \caption{Angular spreading of $|\Omega(\theta,\chi)|^2$ in the $\theta$-direction  for the right-moving mode with $\chi=3.2$.}
        \label{angular_spreading_B=0}
    \end{subfigure}
    \hfill
    \begin{subfigure}{0.45\textwidth}
        \centering
        \includegraphics[width=\linewidth]{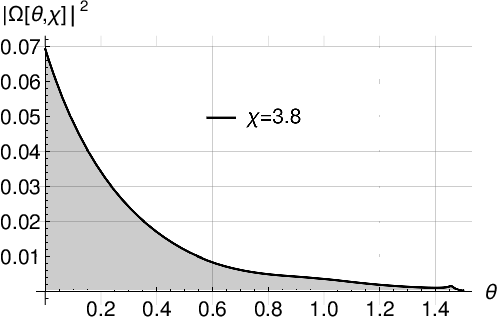}
        \caption{Angular spreading of $|\Omega(\theta,\chi)|^2$ in the $\theta$-direction  for the left-moving mode with $\chi=3.8$.}
        \label{angular_spreading_A=0}
    \end{subfigure}
    \caption{Angular spreading of the string probability density $|\Omega(\theta,\chi)|^2$ in the $\theta$-direction  for right-moving and left-moving modes with constant values of $\chi$.}
    \label{angular spreading}
\end{figure}

Importantly, periodicity in the $\chi$ coordinate implies $\Omega(\theta,\chi)=\Omega(\theta,\chi+2\pi)$, leading to $A(\chi)=A(\chi+2\pi)$ and $B(\chi)=B(\chi+2\pi)$ in equation \ref{eta solution} as $V(y,\chi+2\pi)=V(y,\chi)$. Now because of this choice in equation \ref{1st F}, the exponent $\mathcal{S}=\int^{\theta}\sqrt{V(y,\chi)}dy$ is a complex valued function and hence the solution $\Omega$  in equation \ref{eta solution} is a complex oscillatory function, consisting of two branch. One branch resembles an left-moving wave while the other branch resembles an right-moving wave.

The periodicity of $A(\chi)$ and $B(\chi)$ is consistent with the fact that the constraint \ref{constraint 2} implies $A(\chi)/B(\chi)=\mathbb{F}(\chi)$ is a periodic function of $\chi$ with a period of $2\pi$. Since single valuedness only requires $\mathbb{F}(\chi)$ to be periodic, we may restrict attention to the simplest mode sector in which the ratio $A(\chi)/B(\chi)$ is $\chi$-independent, giving $A(\chi)=\mathcal{K} B(\chi)$. This choice preserves the two-branch structure of the solution and corresponds to a definite mixing of the two independent branches. If one considers a single branch of the string, either $A(\chi)=0$ or $B(\chi)=0$ depending upon which branch is considered.

Having fixed the ratio of the two integration constant, what remains is to fix their functional dependency in $\chi$. 
 Now the periodicity of $A(\chi)$ and $B(\chi)$, one can choose them to have the general form $A(\chi)=C_{+}\exp{[i\mathcal{F}_1(\chi)]}$ and $B(\chi)=C_{-}\exp{[i\mathcal{F}_2(\chi)]}$, where $\mathcal{F}_{1,2}(\chi)$ are some periodic function of $\chi$ with periodicity $2\pi$ with $C_\pm$ being two constants. This choice of the functional dependency ultimately gives the angular probability density $\left|\Omega(\theta,\chi) \right|^2=\frac{C_\pm^2}{(V^*V)^{1/4}}\exp{\{\text{Re}[\pm2\mathcal{S}]\}}$, which is independent of $\mathcal{F}_{1,2}(\chi)$, implying that we can safely choose either $A(\chi)$ or $B(\chi)$ as constants.

To visualize the wave function of the string, we use numerical methods and plot the angular and radial parts of the wave function, choosing the constants as $a_1=1$, $M=10$, $\lambda=1$, $\hbar=1$, $k=2$, $\alpha'=1$, $\beta=1$
in Planck units.

\begin{figure}[h!]
    \centering
    \begin{subfigure}{0.45\textwidth}
        \centering
        \includegraphics[width=\linewidth]{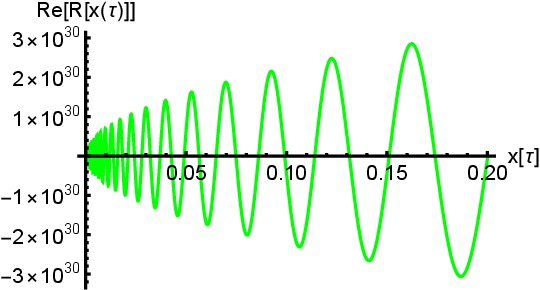}
        \caption{Squeezing in the real part of $R(x)$.}
        \label{angular_spreading_B=0}
    \end{subfigure}
    \hfill
    \begin{subfigure}{0.45\textwidth}
        \centering
        \includegraphics[width=\linewidth]{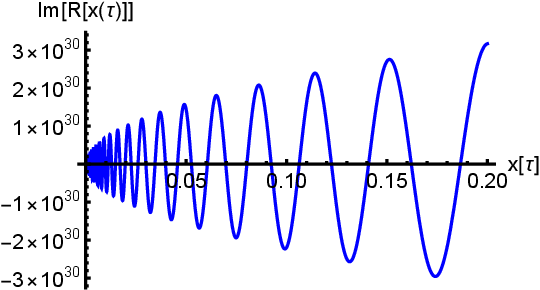}
        \caption{Squeezing in the imaginary part of $R(x)$.}
        \label{angular_spreading_A=0}
    \end{subfigure}
    \caption{Squeezing in the radial sector of string wavefunction $R(x)$ with $M=10$ and $\epsilon=10$ in Planckian units.}
    \label{Radial_squeezing}
\end{figure}

Interestingly, the angular probability density $|\Omega|^2$ in figure \ref{3D eta squared} shows that $|\Omega|^2$ attains peak values in the $\theta$ direction in different regions of the plots depending on the choice of the branch. Fig \ref{eta_squared_B=0}, for the case  $A=1$ and $B=0$,  the density function $|\Omega|^2$ exhibits peaks near the equator $\theta=\pi/2$ of the black hole. On the other hand, for the case $A=0$ and $B=1$, the peak values in $|\Omega|^2$ occur near the pole $\theta=0$.

Although the spherical symmetry is broken, the periodicity of the wavefunction in the $\chi$ coordinate, with period $2\pi$ still remains. This periodicity is imprinted in the metric as well as in the angular sector of the wavefunction in equation \ref{eta solution}. Consequently, the density function of the angular sector also shows this periodicity, as evident from figures \ref{constant_theta_eta_squared_B=0} and \ref{constant_theta_eta_squared_A=0}.

As mentioned earlier, the angular probability density function is not uniform, neither along $\chi$ nor along $\theta$ direction. In the $2$D plot in fig \ref{chi periodicity}, the density function, although shows periodicity, it has maxima as well as minima. On the other hand, it is found in fig \ref{angular spreading} that the density function has either a monotonic increasing or decreasing nature in the $\theta$ direction. Nevertheless, the density function is not uniform in the angular coordinates, suggesting that the probability of finding the string near the vicinity of the horizon depends explicitly on the angles. These are the direct outcomes of the broken spherical symmetry imposed by BMS supertranslation.

The radial part of the wavefunction in figure \ref{Radial_squeezing}  shows the intriguing feature of oscillation with its frequency increasing as the string approaches the horizon while its amplitude decreases. The increase in the frequency is due to gravitational blueshift near the horizon. The decrease in the amplitude of the string near the horizon is a complementary effect of the string increasing in its angular directions. As the string approaches the horizon, it spreads over the angular directions while it gets squeezed along the radial direction, resembling, the phenomenon of string spreading, originally proposed by Susskind \cite{PhysRevD.49.6606} and supported by the findings of \cite{PhysRevD.60.024012}.  Because of the broken $SO(4)$ symmetry, the angular spreading is also non-uniform with maxima and minima, as evident from figure \ref{angular spreading}. It is important to note that although this radial squeezing is independent of the choice of the supertranslation parameter, the angular spreading depends explicitly on $F(\theta,\chi)$. Thus, the near horizon behavior is dominated by gravitational blueshift, while the BMS induced contributions manifest as anisotropies in the angular sector of the wavefunction.

\section{Discussion and Conclusion}\label{disc}
Asymptotic symmetries, particularly those associated with BMS-type transformations, have been argued to play an important role in characterizing the infrared structure of gravitational theories and in encoding information at null and near-horizon boundaries. However, it remains less clear how such symmetry data is reflected in the dynamics of localized or extended quantum probes, especially in higher-dimensional settings where the study of BMS-like transformations is less common. In this regard, extended objects such as strings offer a natural diagnostic, as their worldsheet dynamics couple directly to geometric deformations of the background. This raises the question of whether and how BMS-induced distortions can be detected through the quantum propagation of strings in curved spacetime.

In this work, we therefore investigated the quantum dynamics of a closed bosonic string propagating in the near-horizon geometry of a five-dimensional Schwarzschild black hole subjected to a generalized BMS symmetry transformation. 
Our analysis reveals a distinct decoupling in how the background geometry and the BMS deformations affect the string behaviour. With a legitimate choice of gauge and falloff conditions, the temporal and radial components of the target space metric remain unchanged by the supertranslation. The near-horizon string wavefunction factorizes into temporal, radial, and angular sectors. As a direct effect of the chosen gauge conditions and the structural assumption of the supertranslation parameter $F$, the temporal part of the wavefunction results in a combination of stationary states and the azimuthal part of the wavefunction gives a linear combination of plane wave solution. Furthermore, the periodicity in $\phi$ results in the discretization of its corresponding eigen value $\lambda$.  Consequently, the radial wavefunction is governed by modified Bessel functions, $I_{\pm\sqrt{1-4K}}$, reflecting an effective potential that induces oscillatory behaviour. Crucially, for regimes where $K > 1/4$, we found a non-vanishing, real radial flux defined by $J(x) = \frac{1}{2i}(R^* R' - R R^{*'})$. This non-zero radial current demonstrates that the string configuration corresponds to a transport-type mode, rather than a trapped or purely evanescent state. Furthermore, as the string approaches the horizon, it experiences severe radial squeezing; its amplitude decreases while its frequency increases, an effect driven predominantly by the gravitational blueshift due to the black hole.

Moreover, the imprint of the BMS supertranslations manifests entirely within the angular sector of the string dynamics. The BMS-induced deformation introduces an angle-dependent shear on the three-sphere ($S^3$), explicitly breaking the original $SO(4)$ isometry of the Schwarzschild background and leading to a highly anisotropic geometry. By explicitly modeling the supertranslation parameter $F(\theta, \chi)$ as an $l=2$ hyperspherical mode, we demonstrated that the angular probability density of the string becomes highly non-uniform. Depending on the specific branch of the string wavefunction (incoming versus outgoing), the angular probability density strongly peaks either near the poles or the equator of the black hole.

This dual behaviour---radial squeezing governed by gravity and anisotropic angular spreading governed by the supertranslation---highlights a profound physical consequence, namely,  probe strings can directly detect and physically encode higher-dimensional BMS-like deformations through their worldsheet dynamics. This provides a concrete, dynamical realization of the string spreading paradigm \cite{PhysRevD.49.6606}, where the string spreads angularly over the horizon while being compressed radially. Thus the BMS transformation reorganizes both the spatial distribution and propagation characteristics of the string modes.

While our results provide a foundational step, the current model operates within a restricted framework. We have considered a purely bosonic string in a background spacetime metric,  excluding the effects of Ramond-Ramond (RR) fluxes and worldsheet fermions, to maintain analytical tractability. A natural extension of this work would involve embedding this setup within a fully consistent superstring framework, such as incorporating the Green-Schwarz formalism or the pure spinor formalism to account for fermionic degrees of freedom. Additionally, extending this analysis to rotating backgrounds, such as the $5$D Myers-Perry black hole, would allow exploring the interplay between BMS superrotations, angular momentum, and string spreading. These directions hold the potential to further illuminate the microscopic mechanisms underlying black hole entropy and information storing.

In summary, from a physical perspective, the BMS-induced deformation acts as an  anisotropic distortion of the angular geometry, effectively breaking the  $SO(4)$ symmetry of the undeformed three-sphere. This leads to a lifting  of the degeneracy of hyperspherical modes and induces angular localization  of the string wavefunction. In the radial sector, the near-horizon geometry  gives rise to an effective potential structure, resulting in modified  Bessel-function behaviour, with a non-vanishing radial flux signalling that the corresponding modes are  transport-type configurations. Taken together, these features indicate that probe  strings can encode nontrivial information about BMS-type deformations,  providing a dynamical window into the role of asymptotic symmetries in curved spacetime.

\section{Acknowledgement}
Nihar Ranjan Ghosh is supported through a Research Fellowship from the Ministry of Human Resource Development (MHRD), Government of India.

%\bibliographystyle{IEEEtran}
%\bibliography{paper-07}

\end{document}